# A verification of the Optimal Jet Finder

Fyodor V. Tkachov

INR RAS, Moscow 117312, Russia

The fortran-77 code [hep-ph/9912415] of the jet algorithm based on the optimal jet definition [hep-ph/9901444] has been successfully verified (up to a few interface bugs) using an independently evolved code written in the object-oriented language Component Pascal [http://www.oberon.ch]. The final fortran code OJF_014 is available on the Web.

The optimal jet definition discovered in [1] purports to provide a final theoretical answer to the problem of finding the best jet definition. A practical version of the answer was provided in [2] in the form of a fortran-77 code; its beta (OJF_013) was available on the web at [3].

The first working version of the algorithm was first developed in the object-oriented language Component Pascal using the rapid application development environment BlackBox Component Builder [4]. The BlackBox brings the best features of the Oberon System [5] to commercial environments. Component Pascal is a fine-tuning of the seminal programming language Oberon which has been influencing the software development world in a profound way, and a heir to the famous Pascal and Modula-2 [6]. A superb and unmatched combination of a fast compiler producing high-quality code; strict type safety of the meticulously designed language; and a highly dynamic software development environment allowed to perform the necessary experimentation and find a minimization algorithm to serve as a basis for a fortran implementation which evolved independently to accommodate the fine-tunings of the jet definition (reported in the January 2000 revision of [1]). Large scale tests by Pablo Achard [7] demonstrated usefulness, stability and a good speed of the fortran beta code (OJF_013). However, the following problems were seen:

1) An inferior readability and a lack of safety of fortran made it difficult to make claims about adherence of the encoded algorithm to the theoretical definition of [1].

2) The potential role of the resulting jet algorithm as a universal standard tool in high-energy physics warranted an independent verification of the code.

3) A wide — however unfortunate and deplorable — adoption of C++ as a standard language for software development for high energy physics experiments might at some point require that a C++ port of the algorithm be undertaken. The fortran code does not seem to be the best starting point for that.

To address these problems, it was decided to undertake a complete revision of the early Component Pascal implementation from the ground up without regard to the fortran code and all the optimization implemented therein. The time gap between this undertaking and the switch to fortran in the development of OJF_013, as well as the fact that the fortran code was considerably evolved in the meantime, including a thorough cleanup of the code, ensured as complete independence of the earlier fortran and this last Component Pascal implementations, as is possible given that the two were not done by different people.

The verification version was written in Component Pascal using the BlackBox Component Builder, version 1.4 Educational, and attempted to implement the algorithm as described in [1], [2] in as simple and straightforward manner as possible, avoiding — but providing interface hooks for implementation of — the optimizations which are not essential for mathematical correctness of the algorithm.

Comparison of the fortran beta, OJF_013, with the new Component Pascal code, revealed the following:

(i)  **identity** of the optimal jet configurations found by OJF_013 and the verification version;

(ii)  a minor interface problem in OJF_013 with the set_seed routine (improper handling of large values of the seed, not affecting the correctness of the minimization algorithm itself);

(iii)  a misprint in the formulae for the gradients in [2] (the revision has been posted in the arXive already);

(iv)  a discrepancy in the definition of the function $Y$ between OJF_013 and the main theory paper [1]: the theory paper includes the factor 2 into the definition of Y whereas OJF_013 includes it into the coefficient (Radius2) with which Y is added to E_soft to yield Omega;

(v)  discrepancies due to rounding errors due to slightly different coding of formulas in the two versions; the discrepancies manifest themselves in the form of different local minima sometimes resulting from the same seed in the two versions; this is not a physical problem because both versions find exactly the same global minimum;

(vi)  some superfluous code and intermediate data in OJF_013 — neither affect the results and the functioning of the minimum search algorithm, and their effect on the speed of the algorithm is, so far as I can see, negligible;

(vii)  options for further optimizations.

Taking (i) into account, the version OJF_014 of the fortran code was developed by eliminating (ii)–(iv), which required only very minor changes of the code and did not affect the core minimization routine. The findings (vi) and (vii) were not addressed to avoid spoiling a good thing. The few differences between the beta, OJF_013, and OJF_014 are easily seen using the text comparison feature of text editors.

An additional change in the OJF_014 compared with OJF_013 is a more differential treatment and relaxation of the floating point error bounds used in checks of invariants that must be satisfied at intermediate steps of the algorithm (such as equality to 0 or 1 of certain sums; see the formulae in [2]). These bounds were deliberately chosen very tight in the beta version, causing termination of the algorithm for a small fraction of realistic samples of events. A reasonable relaxation of the bounds cannot break the algorithm as such errors do not propagate from iteration to iteration thanks to the design of the algorithm (especially the mechanism of so-called "snapping" of the recombination matrix elements $z_{aj}$ to the boundary of the simplex). On the whole, no effort was spared to ensure numerical correctness and stability of the code.

The Component Pascal version of the algorithm consists of three independently compiled modules implementing the jet finding algorithm, an I/O module, and a dialog control module. The three main modules implement, respectively:

(1) a minimization algorithm in the $n_\text{jets}$-dimensional standard simplex (recall that the recombination matrix $z_{aj}$ for a fixed particle $a$ is a point in the standard simplex; see [1], [2]) which allows independent checks with arbitrary functions (a new feature compared with the fortran code); this module is implemented in such a way as to allow, say, optimizations similar to those used in the fortran code via redefinition of certain methods without affecting the code of this module;

(2) a module implementing kinematics formulae; polymorphism is used to hide in this module the specifics of the spherical and the hadron-hadron collisions kinematics;

(3) a global minimum search module which allows to vary minimization tactics.

The clean separation of concerns between the modules interacting via explicitly defined interfaces, together with the clean design, excellent readability, and the safety features of Component Pascal minimize chances of introducing bugs so common in fortran and C/C++, so that one would have been inclined to trust the Component Pascal code should a significant difference with the fortran version emerge, which eventuality, thanks goodness, did not materialize.

The resulting fortran code OJF_014 is released as a final public code. It is available from the URL [3].

Still, it is well-known that complex calculations should be done at least twice — independently by different teams. In the case of the optimal jet algorithm, a completely independent implementation of the minimum search algorithm may not be necessary if one notes that the mathematical nature of the global minimum search problem ensures that the results can be rather easily checked in a brute force fashion with OJF_014 treated as a black box. Indeed, it is sufficient to independently write the code evaluating the function $\Omega_R(z_{aj})$ being minimized. Then for a given event, one could generate random configurations of the recombination matrix elements $z_{aj}$ and check that the corresponding values of $\Omega_R$ are not less then the value of $\Omega_R$ on the configuration found by OJF_014. The efficiency of such a verification procedure would be enhanced if one also implements simple procedures to vary the random configurations $z_{aj}$ to decrease $\Omega_R$; such procedures need constitute neither complete, nor efficient minimum search method. It would also be easy e.g. to study in a more detailed fashion the behavior of $\Omega_R$ on the straight line connecting a randomly generated $z_{aj}$ with the point of global minimum produced by OJF_014.

Implementing such a brute force verification should, hopefully, not be a problem with all the hardware out there; one only has to ensure a complete independence of the code for evaluation of $\Omega_R$ from the OJF_014 code.

*Acknowledgments.* This work was supported in parts by the Russian Foundation for Basic Research grant 99-02-18365, and the NATO grant PST.CLG.977751. The sponsorship of the Oberon microsystems, Inc. is gratefully acknowledged.